\documentclass[11pt,twoside]{article}
\usepackage{newpasp}
\usepackage{psfig}
\def\edcomment#1{\iffalse\marginpar{\raggedright\sl#1\/}\else\relax\fi}
\reversemarginpar
\begin{document}
\title{Spectral ageing: a new {\em age} perspective}
\author{Katherine M.\ Blundell and Steve Rawlings}
\affil{Oxford University, Astrophysics, 1 Keble Road, Oxford, OX1 3RH }

\begin{abstract}
We present an up-to-date critique of the physical basis for the
spectral ageing method.  We find that the number of cases where this
method may be meaningfully applied to deduce the ages of classical
double radio sources is small indeed.  This critique is much more than
merely a re-expression of anxieties about the calibration of spectral
ageing (which have been articulated by others in the past).
\end{abstract}

\section{One key observational point}

Many people (e.g.\ Winter et al.\ 1980, Myers \& Spangler 1985,
Alexander \& Leahy 1987) have observed that spectral indices change
along the lobes of classical double radio galaxies.  The general trend
observed is that the lobe spectra are flatter in the outermost regions
near the hotspot and steeper in the regions nearer the core.  Often
the observed change in spectral index, or the spectral gradient, is
steady and systematic.

\subsection{Two {\em a priori} interpretations}

It is not widely acknowledged that there are in principle two physical
interpretations of this behaviour.  The traditional interpretation of
spectral gradients goes as follows: the radiating electrons nearer the
core were dumped by the hotspot much earlier in the past than the
radiating electrons near the hotspot now, and so the former will have
undergone greater synchrotron cooling compared with the latter.  A
radiating population whose energy distribution is initially a power-law,
which suffered only synchrotron losses, would show a `break' in this
power-law at later times.  This break frequency moves to lower frequencies
as more time elapses (Kardashev 1962, Pacholczyk 1970, Jaffe \& Perola
1973) predicting steeper measured spectral indices for the older emission.
Thus far, there is consistency with observations.  But an alternative
physical picture explains the observations just as well: a gradient in
magnetic field along the lobe together with a curved energy electron
spectrum will result in a spectral gradient being observed along the lobe
(see Fig.\,1).  Indeed, Rudnick, Katz-Stone and Anderson's (1994) analysis
of multi-frequency images of Cygnus\,A show no evidence for any variation
in the shape of $N(\gamma)$ across different regions of the lobe.

\begin{figure}
\hbox{
\psfig{figure=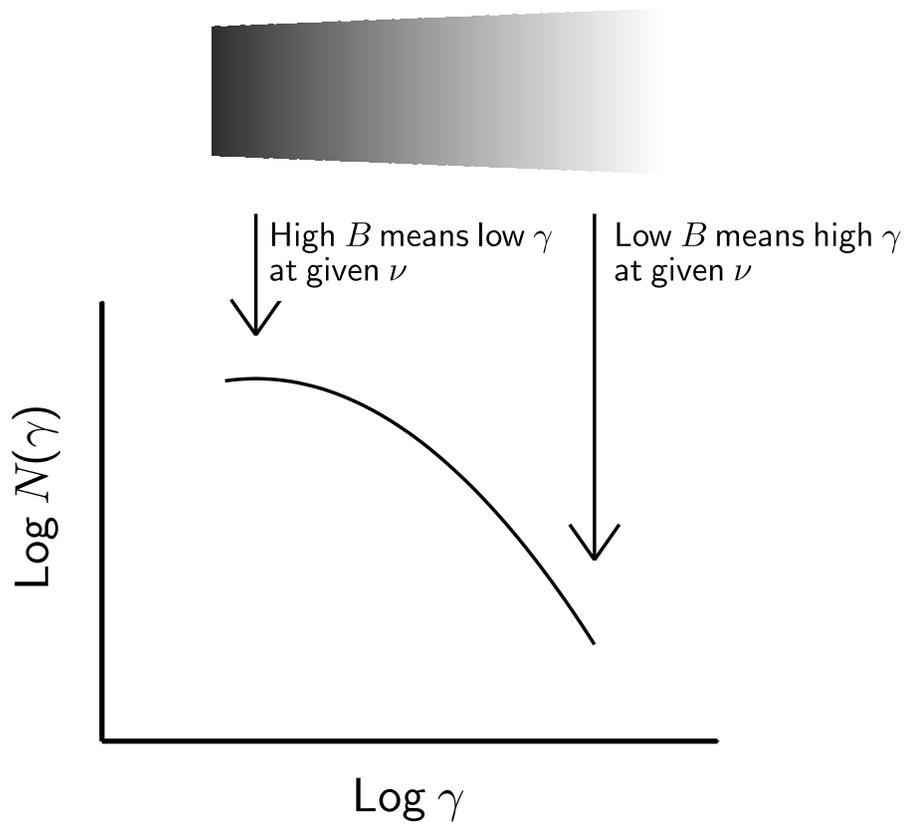,width=0.9\textwidth}
}
\caption{Illustration of how, when one observes a particular frequency
$\nu$ across the lobe, the combination of a gradient in magnetic field
together with a curved electron energy distribution $N(\gamma)$ will
inevitably lead to a gradient in spectral index $\alpha$ ($ \equiv
\partial \log S_{\nu} / \partial \log \nu $).}
\end{figure}

Without considering the underlying physics more deeply, one {\em
cannot} distinguish between these two possibilities.  So we now
examine in turn the individual and collective assumptions which go
into these two pictures.

\section{Assumptions and predictions of the spectral ageing method}

The spectral ageing method usually makes the following assumptions: i) the
magnetic field strength is constant, ii) each segment or slice of lobe may
be regarded as a discrete element of plasma and there is no mixing between
the slices, iii) the radiative lifetimes of the synchrotron particles in
the plasma are significantly longer than the spectral ages to be measured,
and iv) the element of plasma being considered initially has a power-law
distribution in energy.

A `break-frequency' $\nu_{\rm B}$ in this power-law distribution is
therefore deemed to evolve with time according to the following formula:

\begin{equation}
\nu_{\rm B} \propto \frac{1}{B^3t^2}
\end{equation}
where $B$ is the magnetic field strength within the element of plasma and
$t$ is the time which has elapsed since the energy distribution was
accelerated to its deemed power-law distribution.  Sometimes this
expression is slightly modified to incorporate the effects of the
equivalent magnetic field due to inverse Compton scattering off the cosmic
microwave background.

\subsection{Spectral ageing assumption I: constancy of the magnetic
field} The magnetic field strength at the time of observation is
tricky to calibrate: an equipartition field strength has been
frequently used in the spectral ageing methods in the past and such a
value for the magnetic field strength appears to be correct within a
factor of a few (Leahy, these proceedings).

The relationship between the local magnetic field strength $B$ and the
frequency $\nu$ at which most of the energy from particles with
Lorentz factor $\gamma$ will be radiated is given by:

\begin{equation}
\gamma^2 \propto \frac{\nu}{B}.
\end{equation}
Thus, when one considers a particular fixed frequency (e.g.\ 151\,MHz
in the rest frame) observations of sources with {\em lower} magnetic
field strengths will be from synchrotron particles with {\em higher}
Lorentz factors, which are typically from a steeper part of the
underlying energy spectrum.

A recent finding casts considerable doubt on the appropriateness of
assuming that the synchrotron cooling has taken place in a constant
magnetic field strength.  Sources with larger physical sizes have steeper
spectra\footnote{as evaluated at 151\,MHz in the rest frame, hence largely
uncontaminated by core and hotspot emission, and indeed in the frequency
regime whose spectral shape is alleged to be unaffected in the spectral
ageing model, see Fig.\ 2.} (Blundell, Rawlings and Willott 1999).  This
indicates that more expanded sources have lower magnetic field strengths.
Consistently with simple physical pictures of the expansion of radio lobes
(Blundell \& Rawlings 2000) this indicates that the magnetic field
decreases as the lobe expands sideways, as the age of the radio galaxy
increases.

\begin{figure}
\centerline{
\psfig{figure=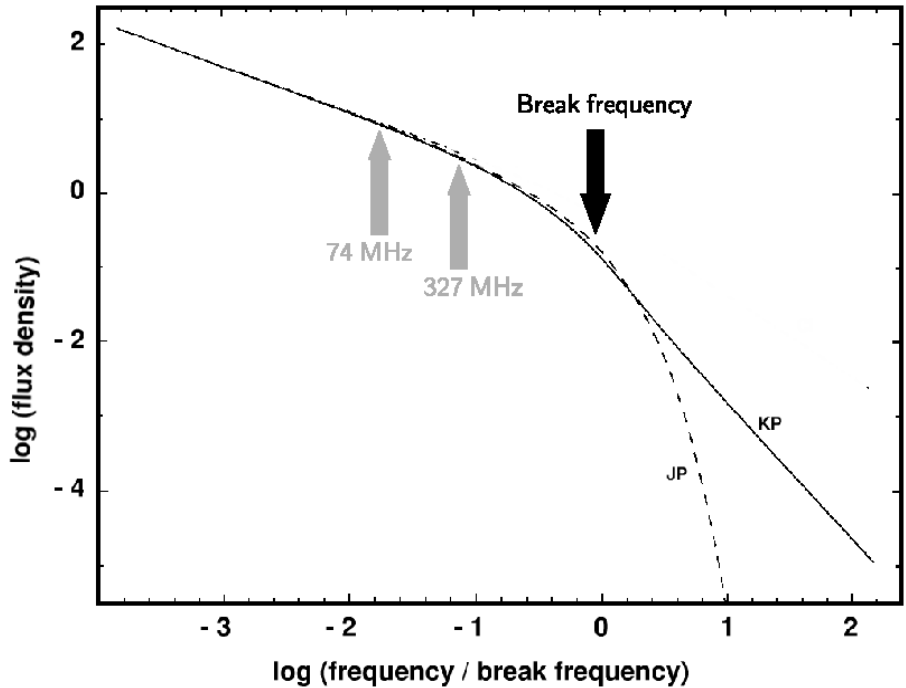,width=0.95\textwidth}
}
\caption{The solid curves are the predicted spectral shapes for two
different models of synchrotron cooling [one with pitch-angle scattering
(Jaffe \& Perola 1973) and one without (Kardashev 1962 and Pacholczyk
1970)]. The large black arrow indicates the `break--frequency' for these
two models.  Carilli et al (1991) fit the break frequencies along the
lobes of Cygnus\,A to be $> 5$\,GHz independent of which of the two models
is used.  Taking the break frequency to be at the lower end of their
measurements i.e.\ 5\,GHz, we plot the frequencies of 74\,MHz and
330\,MHz. Inconsistently with the above picture, between these frequencies
a spectral index gradient is actually observed.  This beckons an
alternative physical picture to that assumed in the spectral ageing
method, see Fig.\ 1.}
\end{figure}

\subsection{Spectral ageing assumption II: no mixing of populations}  
We now consider spectral ageing input assumption II.  Jones, Ryu and
Engel (1999) pointed out that any mixing of particles within the lobes
will contaminate spectral ageing estimates.  Possible origins of
mixing include: bulk backflow and turbulent backflow as well as a
variety of diffusion mechanisms.  Which of these is the most effective
transport mechanism depends on the details of the magnetic field
configuration within the lobe; assuming a diffusion rate of zero is
perhaps taking a liberty.

\subsection{Spectral ageing assumption III: input/hotspot spectra are
power-laws}
We now consider spectral ageing input assumption III that the initial
injection index is a straight power-law.  State-of-the-art
observations of hotspots over a wide frequency baseline show that
their spectra are {\em curved} (see e.g.\ Carilli et al.\ 1991).
Since the hotspots are believed to supply plasma into the lobes, it is
hard to see how a curved input spectrum would be able to straighten
itself out once in the lobe.

Moreover, the observed low-frequency asymptote is inconsistent with that
assumed in the spectral ageing method: spectral gradients are observed
well below the break frequencies in some lobes (see Fig.\ 2).  The new
74\,MHz receiver system on the VLA has enabled spatially resolved images
at super-metre wavelengths for the first time.  The spectral gradient in
Cygnus A between 74\,MHz and 330\,MHz first published by Kassim et al
(1996), shows a clear spectral gradient at frequencies well below the
break frequencies fitted by Carilli et al (1991).  This appears to be
unexceptional behaviour for classical double radio galaxies (Perley et
al.\ in preparation).  This behaviour would not be observed if the
spectral shape at these frequencies were power-laws.  This observation is
more consistent with assuming that there is a magnetic field gradient
along the lobe.

\subsection{Spectral ageing assumption IV: the radiative lifetimes are
long enough}  

The spectral ageing model (e.g.\ Fig.\,2) assumes that the radiative
lifetimes of the synchrotron emitting particles (or the `cooling
times') are longer than the ages which are to be deduced.  Matthews
and Scheuer (1990) were the first to consider the evolution of the
energies of the synchrotron particles, rather than the more commonly
used evolution of the frequency spectrum.  Their expression gives
${\rm d}\gamma/{\rm d} t$ if both synchrotron and adiabatic expansion
losses are occurring:
\begin{equation}
-\frac{{\rm d}\gamma}{{\rm d} t} = \frac{2}{3}{\sigma_{\rm
 T}}{mc\mu_0} \gamma^2 b^2 + \frac{\gamma}{R}\frac{{\rm
 d}R}{{\rm d}t}.
\end{equation}
A similar relation, which includes inverse Compton losses off the cosmic
microwave background, is given by Kaiser, Dennett-Thorpe and Alexander
(1997).

If we consider some particular observing frequency, we can define a
maximum time $\Delta t$ which can elapse between an initial time
$t_{\rm min}$ when a particle is injected into the lobe with a
particular Lorentz factor $\gamma$ and the time of observation if the
particle is to emit synchrotron radiation at the {\em specified
observing frequency} at the time of observation.  This $\Delta t$ can
then be compared with the age of the radio source.  Prior to this time
$t_{\rm min}$, even if particles of extremely high Lorentz factor are
injected into the lobe, their enhanced energy losses will be so
catastrophic that their Lorentz factors will be too low at the time of
observation\footnote{We use the term `time of observation' quite
liberally here to mean `when the source is intercepted by our
light-cone' or `when the light we ultimately observe leaves the
source'.} ($t_{\rm obs}$) to contribute to radiation at the given
frequency, in a given $B$-field.

Thus for the ensemble of particles contributing to the radiation
at a given emitted frequency at $t_{\rm obs}$, those which had
the largest Lorentz factors at injection are injected at $t_{\rm
min}$ and those with the lowest are those actually injected at
$t_{\rm obs}$.  There is no trivial identity which connects
$t_{\rm min}$ with the age of the source, as Fig.\ 3
shows.

\begin{figure}
\null\hfill
\hbox{
\psfig{figure=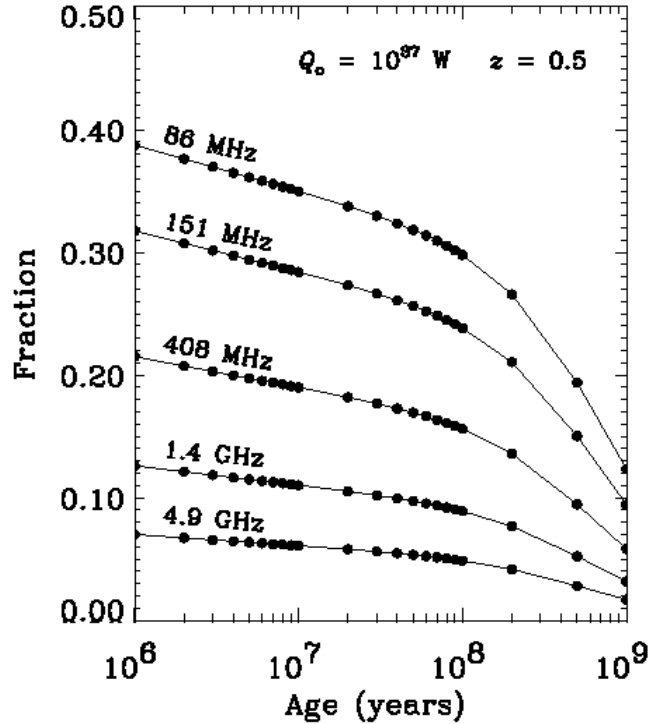,width=0.65\textwidth}
}
\hfill\null
\caption{This plot shows the fraction of the age of the radio source for
which the particles radiating at a certain frequency have been in the
radio lobe, against the age of the radio-source.  The bulk kinetic
jet-power in the model source shown in this plot is $10^{37}\,{\rm W}$
(for details of the assumed environment see Blundell \& Rawlings 2000.)
In this model the magnetic field gently decreases with time as the lobes
expand.  Even at low frequencies ($< 200\,{\rm MHz}$) the synchrotron
lifetime of the radiating particles is significantly shorter than the
current age of the source.  }
\end{figure}

\subsection{Prediction of spectral ageing I}
A prediction of spectral ageing is that the derived spectral ages should
be consistent with the dynamical ages.  The simplest constraints on source
ages come from the measured projected physical size of an object and
estimating the speed at which it has expanded to that size.  The
dimensional model of Falle (1991) has that the expansion speed of a radio
galaxy {\em decreases} as the source gets {\em older} (in a given
environment), so although proper motion measurements with VLBI techniques
show that the smallest double radio sources appear to expand at $\sim
0.2\,c$ (Owsianik \& Conway 1998, Owsianik, Conway and Polatidis 1998)
this does not much constrain the expansion speeds of the large classical
double radio sources.  A more promising constraint comes from
considerations of lobe-length asymmetries based on light-travel time
arguments.  These arguments benefit considerably from knowing which lobe
is the nearer to us.  Peter Scheuer, to whose memory this volume is
dedicated, in 1995 used the presence of a jet in FR\,II quasars to infer
speeds of $< 0.05c$.  Arshakian (these proceedings) inferred slightly
larger velocities for a somewhat different sample of FR\,II radio galaxies
and quasars.  Scheuer's speeds imply ages which are typically an
order of magnitude larger than the spectral ages (Alexander \& Leahy
1987).

\subsection{Prediction of spectral ageing II}
It was pointed out a number of decades ago (Jenkins \& Scheuer 1976) that
{\em if} synchrotron cooling played a part in the spectral shape of
extended lobes, then the lobes should be more extended at lower
frequencies.  This rarely appears to be the case.  For example, the
observations by Leahy, Muxlow \& Stephens (1989) of some 3C sources at
151\,MHz with MERLIN and the VLA at 1.4\,GHz show that the lobe lengths at
these different frequencies are the same.  (The flux-density measured by
MERLIN is consistent with that measured by low-resolution instruments.)

\subsection{Prediction of spectral ageing III} 
If the assumptions used in the spectral ageing model were correct,
then there should be many more relic radio galaxies observed, i.e.\
there should be no prohibition on seeing hotspot-less, core-less lobes
in the low-frequency sky.  Such relic radio galaxies are however very
rare with only a very few examples known (e.g.\ Cordey 1987).  This
strongly indicates that the radiative lifetimes of synchrotron
particles in the lobes are not the three orders of magnitude larger
than those of particles in hotspots required by the spectral ageing
method.

\section{Conclusion}
The traditional way of interpreting observed gradients in spectral index
along the radio lobes is the simple spectral ageing picture; we find that
the assumptions which underlie this model, both individually and
collectively, are flawed.

A combination of a gradient in magnetic field (which is physically
plausible) together with a curved distribution in electron energy
(which is measured) produces the same observed behaviour.  In
addition, this second model explains not just the observed spectral
gradients {\em above} the break frequency but also those {\em below}
the break frequencies.


\begin{references}
\reference Alexander, P.\ \& Leahy, J.P., 1987, `Ageing and speeds in
a representative sample of 21 classical double radio sources', 
\mnras, 225, 1-26

\reference Blundell, K.M.\ \& Rawlings, S., 2000, 
`The spectra and energies of classical double radio lobes',
\aj, 119, 1111-1122

\reference Blundell, K.M., Rawlings, S.\ \& Willott, C.J., 1999,
`The nature and evolution of classical double radio sources from
complete samples',
\aj, 117, 667-706

\reference Carilli, C.L., Perley, R.A., Dreher, J.W., \& Leahy, J.P.,
1991, 
`Multifrequency radio observations of Cygnus A --- spectral aging in
powerful radio galaxies',
\apj, 383, 554-573

\reference Cordey, R.A., 1987, 
`IC 2476 - A possible relic radio galaxy',
\mnras, 227, 695-700

\reference Falle, S.A.E.G., 1991, 
`Self-similar jets',
\mnras, 250, 581-596

\reference Jenkins, C.R.\ \& Scheuer, P.A.G., 1976, 
`What docks the tails of radio source components?',
\mnras, 174, 327-333

\reference Jones, T.W., Ryu, D.\ \& Engel, A., 1999, 
`Simulating electron transport and synchrotron emission in radio galaxies: 
shock acceleration and synchrotron aging in axisymmetric flows',
\apj, 512, 105-124

\reference Kaiser, C.R., Dennett-Thorpe, J., \& Alexander, P., 1997,
`Evolutionary tracks of FRII sources through the P-D diagram',
\mnras, 292, 723-732

\reference Kardashev, N.S., 1962, 
`Nonstationariness of spectra of young sources of nonthermal radio
emission',
SvA, 6, 317-327

\reference Kassim, N.E., Perley, R.A., Carilli, C.L., Harris, D.E.\ \&
Erickson, W.C., 1996, 
`Low frequency observations of Cygnus A',
in `Cygnus A -- Study of
a radio galaxy', 182-190, eds C.L.\ Carilli \& D.E.\ Harris

\reference Leahy, J.P., Muxlow, T.W.B., \& Stephens, P.W., 1989,
`151-MHz and 1.5-GHz observations of bridges in powerful extragalactic
radio sources',
\mnras, 239, 401-440

\reference Matthews, A.P.\ \& Scheuer, P.A.G., 1990, 
`Models of radio galaxies with tangled magnetic fields. I -- calculation
of magnetic field transport, Stokes parameters and synchrotron losses',
\mnras, 242, 616-635

\reference Myers, S.T.\ \& Spangler, S.R., 1985, 
`Synchrotron aging in the lobes of luminous radio galaxies',
\apj, 291, 52-62

\reference Owsianik, I.\ \& Conway, J.E., 1998, 
`First detection of hotspot advance in a Compact Symmetric Object. Evidence
for a class of very young extragalactic radio sources',
\aap, 337, 69-79

\reference Owsianik, I., Conway, J.E.\ \& Polatidis, A.G., 1998, 
`Renewed radio activity of age 370 years in the extragalactic source
0108+388',
\aap, 336, L37-L40

\reference Pacholczyk, A.G., 1970, `Radio Astrophysics', pub.\
W. H. Freeman \& Co 

\reference Jaffe, W.J.\ \& Perola, G.C.\ 1973, 
`Dynamical models of tailed radio sources in clusters of galaxies',
\aap, 26, 423--435

\reference Rudnick, L., Katz-Stone, D.M.\ \& Anderson, M.C., 1994,
`Do relativistic electrons either gain or lose energy, outside of
extragalactic nuclei?',
ApJS, 90, 955--958

\reference Scheuer, P.A.G., 1995, 
`Lobe Asymmetry and the Expansion Speeds of Radio Sources', 
\mnras, 277, 331--340

\reference Winter, A.J.B., Wilson, D.M.A., Warner, P.J., Waldram, E.M.,
Routledge, D., Nicol, A.T., Boysen, R.C., Bly, D.W.J.\ \&
Baldwin, J.E., 1980, 
`The structure of Cygnus A at 150 MHz',
\mnras, 192, 931--944
\end{references}
\end{document}